\documentclass[prd,twocolumn,floatfix,amsmath,nofootinbib,amssymb,floatfix]{revtex4}
\usepackage{graphicx,color,dcolumn,booktabs,bm}
\usepackage{longtable,lscape}
\usepackage{pdfpages}
\usepackage{txfonts}
\usepackage{overpic}
\usepackage{amssymb}
\usepackage{makecell}
\usepackage{indentfirst}
\usepackage{feynmf}   
\usepackage{slashed}  
\usepackage{cases}
\usepackage{color}
\usepackage{multirow}
\usepackage{threeparttable}
\usepackage{epstopdf}
\usepackage{enumerate}
\usepackage{diagbox}
\usepackage{graphicx,color,dcolumn,booktabs,bm}
\usepackage[colorlinks,
            citecolor=blue,
            anchorcolor=red,
            menucolor=red,
            linkcolor=red,
            filecolor=red,
            runcolor=red,
            urlcolor=blue,
            frenchlinks=red]{hyperref}
\usepackage{tikz}

\begin{document}

\title{Molecular-type $QQss\bar{s}$ pentaquarks predicted by an extended hidden gauge symmetry approach}

\author{Zhong-Yu Wang$^{1,2}$}
\email{zhongyuwang@foxmail.com}

\author{Chu-Wen Xiao$^{3,4,5}$}
\email{xiaochw@gxnu.edu.cn}

\author{Zhi-Feng Sun$^{1,2,6,7}$}
\email{sunzf@lzu.edu.cn}

\author{Xiang Liu$^{1,2,6,7}$}
\email{xiangliu@lzu.edu.cn}

\affiliation{
$^1$School of Physical Science and Technology, Lanzhou University, Lanzhou 730000, China \\
$^2$Lanzhou Center for Theoretical Physics, Key Laboratory of Theoretical Physics of Gansu Province, and Key Laboratory of Quantum Theory and Applications of the Ministry of Education, Lanzhou University, Lanzhou, 730000, China \\
$^3$Department of Physics, Guangxi Normal University, Guilin 541004, China \\
$^4$Guangxi Key Laboratory of Nuclear Physics and Technology,
Guangxi Normal University, Guilin 541004, China \\
$^5$School of Physics, Central South University, Changsha 410083, China \\
$^6$MoE Frontiers Science Center for Rare Isotopes, Lanzhou University, Lanzhou 730000, China \\
$^7$Research Center for Hadron and CSR Physics, Lanzhou University and Institute of Modern Physics of CAS, Lanzhou 730000, China
}

\date{\today}

\begin{abstract}
In this work, we investigate the double-heavy  molecular pentaquark states with the quark contents $ccss\bar{s}$, $bbss\bar{s}$, and $bcss\bar{s}$ by using the coupled channel approach. 
The extended local hidden gauge Lagrangians are used to obtain the meson-baryon interactions by exchanging the vector mesons. 
We predict some candidates for the molecular states with the quantum numbers $I(J^{P}) = 0(1/2^{-}, 3/2^{-}, 5/2^{-})$, whose binding energies are of the order of $20-30$ MeV and whose widths are all less than $8$ MeV.
These predicted exotic double-heavy molecular pentaquark states may be accessible in future experiments such as LHCb.
\end{abstract}
\maketitle

\section{Introduction}\label{sec:Introduction}

How to quantitatively describe the strong interaction at the low energy region is a challenging problem in particle physics. Hadron spectroscopy can provide some insights to deepen our understanding of the non-perturbative behavior of the strong interaction. 
Usually, there are  conventional mesons and baryons, which have the quark components of $q\bar{q}$ and $qqq$, respectively. These conventional hadrons form the bulk of the reported hadrons, as shown in the Particle Data Group (PDG) \cite{ParticleDataGroup:2022pth}. It is the first stage in the construction of the particle zoo.

Since 2003, we have entered a new phase. More and more heavy flavor new hadronic states have been found in experiments, which cannot be explained by the quenched quark model. For example, there were famous low-mass puzzles for these reported $D_{s0}^{*}(2317)$ \cite{BaBar:2003oey}, $D_{s1}(2460)$ \cite{CLEO:2003ggt}, and $X(3872)$ \cite{Belle:2003nnu}. 
In 2015, the LHCb Collaboration discovered the pentaquark-like states $P_{c}(4380)^{+}$ and $P_{c}(4450)^{+}$ in the decay of $\Lambda_{b}^{0}\rightarrow J/\psi K^{-}p$ \cite{LHCb:2015yax}, which were considered as competing candidates for the hadronic molecular states and attracted intense interest in both theory and experiment.
In 2017, the observation of the doubly charmed baryon $\Xi_{cc}^{++}$ in the $\Lambda_{c}^{+}K^{-}\pi^{+}\pi^{+}$ invariant mass spectrum by the LHCb Collaboration \cite{LHCb:2017iph} captured the continuing enthusiasm for the heavy states.
Furthermore, with the discovery of several narrow states $\Omega_{c}$ \cite{LHCb:2017uwr,Belle:2017ext,LHCb:2021ptx,LHCb:2023rtu} and $\Omega_{b}$ \cite{LHCb:2020tqd,Matiunin:2020xbg}, the LHCb Collaboration tried to search for more doubly heavy baryons, such as $\Xi_{bc}$, $\Omega_{cc}$, and $\Omega_{bc}$, etc \cite{LHCb:2021xba}. More results can be found in these recent review articles \cite{Liu:2013waa,Hosaka:2016pey,Chen:2016qju,Richard:2016eis,Lebed:2016hpi,Olsen:2017bmm,Guo:2017jvc,Liu:2019zoy,Brambilla:2019esw,Meng:2022ozq,Chen:2022asf}. 
These new hadronic states not only provide a good opportunity to continue building the particle zoo, but also make it possible to search for exotic states. 

In fact, the study of heavy flavor pentaquarks has always been a focus point in the search for exotic states, as reflected in the recent observations of $P_{cs}(4338)^{0}$ and $P_{cs}(4459)^{0}$, which are from the $B^-\to J/\psi \Lambda \bar{p}$ and $\Xi_b^-\to J/\psi \Lambda K^{-}$ processes, respectively \cite{LHCb:2022ogu,LHCb:2020jpq}. The observation of the characteristic spectrum of $P_c(4312)^{+}$, $P_c(4440)^{+}$, and $P_{c}(4457)^{+}$ \cite{LHCb:2015yax,LHCb:2019kea} supports the existence of the molecular-type hidden-charm pentaquark \cite{Wu:2010jy,Wang:2011rga,Yang:2011wz,Li:2014gra,Chen:2015loa,Karliner:2015ina}, and other such states were predicted in Refs. \cite{Chen:2015moa,Ozdem:2018qeh,Du:2019pij,Azizi:2022qll}. 
The discovered $T_{cc}(3875)$ \cite{LHCb:2021vvq} is a good candidate for a double-charm tetraquark \cite{Manohar:1992nd,Ericson:1993wy,Tornqvist:1993ng,Janc:2004qn,Ding:2009vj,Molina:2010tx,Ohkoda:2012hv,Li:2012ss,Xu:2017tsr,Liu:2019stu,Tang:2019nwv,Ding:2020dio}. 
These phenomena mentioned above give us sufficient reason to believe that there should be double-charm molecular pentaquarks \cite{Hofmann:2005sw,Romanets:2012hm,Zhou:2018bkn,Dong:2021bvy,Wang:2022aga,Wang:2023aob}, which is the starting point of the present work. 

In this work, we choose a special double-heavy molecular pentaquark system. This system contains most strange quarks. For quantitatively obtaining the information of the mass spectrum of this focused pentaquark system, we adopt an extension of the chiral unitary approach (ChUA) based on the local hidden gauge approach. 
The ChUA was proposed early on to describe the states $f_{0}(500)$, $f_{0}(980)$, $a_{0}(980)$, and $\Lambda(1405)$ in the molecular configuration \cite{Kaiser:1995eg,Oller:1997ti,Oset:1997it,Kaiser:1998fi,Oller:1998hw,Oller:2000ma,Oller:2000fj}. Note that the scattering amplitudes of the coupled channels were calculated by solving the Bethe-Salpeter equation using the on-shell approximation in Refs. \cite{Oller:1997ti,Oset:1997it,Oller:2000fj}.
This approach has been widely applied in the meson-meson and meson-baryon interactions, with great success, for example in the predictions of the $P_{c}$, $P_{cs}$ states \cite{Wu:2010jy,Wu:2010vk,Xiao:2013yca,Xiao:2019gjd} and the interpretation of the $T_{cc}$ state \cite{Feijoo:2021ppq,Dai:2023cyo}. 
Using this approach, Ref. \cite{Debastiani:2017ewu} dynamically reproduced three $\Omega_{c}$ states observed in the $\Xi_{c}^{+}K^{-}$ invariant mass spectrum by the LHCb Collaboration \cite{LHCb:2017uwr}. 
Moreover, taking the advantage of this method, it has been used to predict the molecular states in the heavy sector, such as the resonances $\Omega_{b}$ \cite{Liang:2017ejq,Liang:2020dxr}, $\Xi_{cc}$ \cite{Dias:2018qhp}, $\Xi_{c}$ and $\Xi_{b}$ \cite{Yu:2018yxl}, $\Xi_{bc}$ \cite{Yu:2019yfr}, $\Xi_{bb}$ and $\Omega_{bbb}$ \cite{Dias:2019klk}, and so on.
Note that the systems with the quark components $ccsq\bar{q}$, $bbsq\bar{q}$, and $bcsq\bar{q}$ were studied in the work of \cite{Wang:2022aga}, where some states of $\Omega_{cc}$, $\Omega_{bb}$, $\Omega_{bc}$ were predicted. 

In order to provide more theoretical support to experiment, in the present work we study the double-heavy molecular pentaquark systems with the quark components $ccss\bar{s}$, $bbss\bar{s}$, and $bcss\bar{s}$, which are dynamically generated from the $s$-wave interactions of meson and baryon using an extension of the ChUA based on the local hidden gauge approach. 
For the meson-baryon interactions, the dominant potentials are obtained by the vector meson exchange mechanism from the extended local hidden gauge Lagrangians as done in Ref. \cite{Debastiani:2017ewu}.
In fact, the diagonal terms of the matrix for the interaction potentials exchange only the light vector mesons, and thus in the interaction procedures the heavy quarks act as spectators, which automatically respects the heavy quark symmetry. 
On the other hand, the non-diagonal terms that exchange the heavy vector mesons do not fulfil the heavy quark symmetry, but neither should they since they correspond to subleading terms in the heavy quark counting, see further discussion in Ref. \cite{Debastiani:2017ewu}. 

The rest of the paper is organized as follows. 
The framework for the potentials of meson-baryon interactions derived from the local hidden gauge Lagrangians is introduced in Sec. \ref{sec:Formalism}. 
The numerical results of the scattering amplitudes solved by the coupled channel Bethe-Salpeter equation are presented in Sec. \ref{sec:Results}. Finally, we conclude this work with a summary in Section \ref{sec:Conclusions}.

\section{The molecular-type double-heavy pentaquarks with three strange quarks}\label{sec:Formalism}

In this section, we will briefly introduce the formalism. First, following Ref. \cite{Wang:2022aga}, we construct the coupled channels of the systems with the quark components $ccss\bar{s}$, $bbss\bar{s}$, and $bcss\bar{s}$. 

In the $ccss\bar{s}$ sector, We divide the coupled channels into four blocks, $PB(1/2^{+})$, $PB(3/2^{+})$, $VB(1/2^{+})$, and $VB(3/2^{+})$, as shown in Table \ref{tab:coupledchannels}, where $P$ stands for the pseudoscalar meson, $V$ for the vector meson, $B(1/2^{+})$ for the $J^{P}=1/2^{+}$ ground state baryons, and $B(3/2^{+})$ for the $J^{P}=3/2^{+}$ ground state baryons. 
Note that there is no channel containing the $\eta$ meson as well as the quark components $ccsq\bar{q}$ $(q=u,d)$, as these systems were studied in Ref. \cite{Wang:2022aga}, and the thresholds are much lower than those of the present work.
A similar situation occurs in the $bbss\bar{s}$ and $bcss\bar{s}$ sectors.

\begin{table}[htbp]
\centering
\renewcommand\tabcolsep{4.5mm}
\renewcommand{\arraystretch}{1.50}
\caption{Coupled channels in different sectors. The symbol $P$ stands for the pseudoscalar meson, $V$ for the vector meson, $B(1/2^{+})$ for the $J^{P}=1/2^{+}$ ground state baryons, and $B(3/2^{+})$ for the $J^{P}=3/2^{+}$ ground state baryons.}
\begin{tabular*}{86mm}{l|cccc}
\toprule[1.00pt]
\toprule[1.00pt]
&\multicolumn{2}{c|}{\mbox{$ccss\bar{s}$}}&\multicolumn{2}{c}{\mbox{$bbss\bar{s}$}}\\
\hline
$PB(1/2^{+})$&$\eta^{'}\Omega_{cc}$&\multicolumn{1}{c|}{$D_{s}\Omega_{c}$}&$\eta^{'}\Omega_{bb}$&$\bar{B}_{s}\Omega_{b}$\\
$PB(3/2^{+})$&$\eta^{'}\Omega_{cc}^{*}$&\multicolumn{1}{c|}{$D_{s}\Omega_{c}^{*}$}&$\eta^{'}\Omega_{bb}^{*}$&$\bar{B}_{s}\Omega_{b}^{*}$\\
$VB(1/2^{+})$&$\phi\Omega_{cc}$&\multicolumn{1}{c|}{$D_{s}^{*}\Omega_{c}$}&$\phi\Omega_{bb}$&$\bar{B}_{s}^{*}\Omega_{b}$\\
$VB(3/2^{+})$&$\phi\Omega_{cc}^{*}$&\multicolumn{1}{c|}{$D_{s}^{*}\Omega_{c}^{*}$}&$\phi\Omega_{bb}^{*}$&$\bar{B}_{s}^{*}\Omega_{b}^{*}$\\
\hline
\multicolumn{1}{c}{}&\multicolumn{4}{c}{\mbox{$bcss\bar{s}$}}\\
\hline
$PB(1/2^{+})$&$\eta^{'}\Omega_{bc}$&$\eta^{'}\Omega_{bc}^{'}$&$D_{s}\Omega_{b}$&$\bar{B}_{s}\Omega_{c}$\\
$PB(3/2^{+})$&$\eta^{'}\Omega_{bc}^{*}$&$D_{s}\Omega_{b}^{*}$&$\bar{B}_{s}\Omega_{c}^{*}$&\\
$VB(1/2^{+})$&$\phi\Omega_{bc}$&$\phi\Omega_{bc}^{'}$&$D_{s}^{*}\Omega_{b}$&$\bar{B}_{s}^{*}\Omega_{c}$\\
$VB(3/2^{+})$&$\phi\Omega_{bc}^{*}$&$D_{s}^{*}\Omega_{b}^{*}$&$\bar{B}_{s}^{*}\Omega_{c}^{*}$&\\
\bottomrule[1.00pt]
\bottomrule[1.00pt]
\end{tabular*}
\label{tab:coupledchannels}
\end{table}

The evaluation of the potentials for the meson-baryon interactions is given by the extended local hidden gauge symmetry approach \cite{Wang:2022aga}. 
The interaction mechanism in each block is constructed from the dominant process of the vector meson exchange.
The local hidden gauge Lagrangians \cite{Bando:1984ej,Bando:1987br,Meissner:1987ge} describing the vertices $VPP$, $VVV$, and $VBB$ are given by
\begin{equation}
\begin{aligned} 
\mathcal{L}_{V P P}=-i g\left\langle\left[P, \partial_\mu P\right] V^\mu\right\rangle,
\end{aligned}
\label{eq:LVVP}
\end{equation}
\begin{equation}
\begin{aligned} 
\mathcal{L}_{V V V}=i g\left\langle\left(V^\mu \partial_\nu V_\mu-\partial_\nu V^\mu V_\mu\right) V^\nu\right\rangle,
\end{aligned}
\label{eq:LVVV}
\end{equation}
\begin{equation}
\begin{aligned}
\mathcal{L}_{V B B}=g\left(\left\langle\bar{B} \gamma_\mu\left[V^\mu, B\right]\right\rangle+\left\langle\bar{B} \gamma_\mu B\right\rangle\left\langle V^\mu\right\rangle\right),
\end{aligned}
\label{eq:LVBB}
\end{equation}
where $g$ is the coupling constant, which is taken as $M_{V}/(2f_{\pi})$ with $M_{V}$ the mass of the exchanged light vector meson and $f_{\pi}=93$ MeV the pion decay constant. The symbol $\left\langle\right\rangle$ denotes the trace of flavor space. 
Note that Eq. \eqref{eq:LVBB} only holds for SU(3), light baryons, which is not used in our case, see the discussions later.
For the pseudoscalar $P$ and vector $V^{\mu}$ meson fields, they are easily extended from the SU(3) to the SU(5) case following the work of \cite{Hofmann:2005sw,Mizutani:2006vq},
\begin{equation}
\begin{aligned}
P=\left(\begin{array}{ccccc}
\frac{\eta}{\sqrt{3}}+\frac{\eta^{\prime}}{\sqrt{6}}+\frac{\pi^0}{\sqrt{2}} & \pi^{+} & K^{+} & \bar{D}^0 & B^{+} \\
\pi^{-} & \frac{\eta}{\sqrt{3}}+\frac{\eta^{\prime}}{\sqrt{6}}-\frac{\pi^0}{\sqrt{2}} & K^0 & D^{-} & B^0 \\
K^{-} & \bar{K}^0 & -\frac{\eta}{\sqrt{3}}+\sqrt{\frac{2}{3}}\eta^{\prime} & D_s^{-} & B_s^0 \\
D^0 & D^{+} & D_s^{+} & \eta_c & B_c^{+} \\
B^{-} & \bar{B}^0 & \bar{B}_s^0 & B_c^{-} & \eta_b
\end{array}\right),
\end{aligned}
\label{eq:P}
\end{equation}
\begin{equation}
\begin{aligned}
V^{\mu}=\left(\begin{array}{ccccc}
\frac{\omega+\rho^0}{\sqrt{2}} & \rho^{+} & K^{*+} & \bar{D}^{* 0} & B^{*+} \\
\rho^{-} & \frac{\omega-\rho^0}{\sqrt{2}} & K^{* 0} & D^{*-} & B^{* 0} \\
K^{*-} & \bar{K}^{* 0} & \phi & D_s^{*-} & B_s^{* 0} \\
D^{* 0} & D^{*+} & D_s^{*+} & J / \psi & B_c^{*+} \\
B^{*-} & \bar{B}^{* 0} & \bar{B}_s^{* 0} & B_c^{*-} & \Upsilon
\end{array}\right)^{\mu},
\end{aligned}
\label{eq:V}
\end{equation}
but one is not making use of SU(5) symmetry, only of the $q \bar{q}$ character of the mesons, as shown in Ref. \cite{Sakai:2017avl}.

We can easily obtain the $VPP$ and $VVV$ vertices in the charm and bottom sectors from Eqs. (\ref{eq:LVVP})-(\ref{eq:LVVV}) and (\ref{eq:P})-(\ref{eq:V}). 
However, it cannot directly extend the SU(3) $VBB$ vertex to the charm and bottom sectors. According to Ref. \cite{Debastiani:2017ewu}, the $VBB$ vertex can be constructed through the flavor and spin wave functions of mesons and baryons. 
In Table \ref{tab:Wavefunctions}, we list the flavor and spin wave functions of baryons used in this work, following the convention of Ref. \cite{Close:1979}. 
Thus, for the $J^{P}=1/2^{+}$ ground state baryons in the case of $S_{z}=+1/2$ we have

\begin{table}[htbp]
\centering
\renewcommand\tabcolsep{2mm}
\renewcommand{\arraystretch}{1.50}
\caption{Wave functions for baryons with $I(J^{P})=0(\frac{1}{2}^{+})$ and $0(\frac{3}{2}^{+})$, where $\chi_{MS}$ stands for mixed symmetric, $\chi_{MA}$ stands for mixed antisymmetric, and $\chi_{S}$ stands for fully symmetric.}
\begin{tabular*}{86mm}{@{\extracolsep{\fill}}lccc}
\toprule[1.00pt]
\toprule[1.00pt]
States&$I(J^{P})$&Flavor& Spin\\
\hline
$\Omega_{cc}^{+}$&$0(\frac{1}{2}^{+})$&$ccs$&$\chi_{MS}(12)$\\
$\Omega_{c}^{0}$&$0(\frac{1}{2}^{+})$&$css$&$\chi_{MS}(23)$\\
$\Omega_{bb}^{-}$&$0(\frac{1}{2}^{+})$&$bbs$&$\chi_{MS}(12)$\\
$\Omega_{b}^{-}$&$0(\frac{1}{2}^{+})$&$bss$&$\chi_{MS}(23)$\\
$\Omega_{bc}^{0}$&$0(\frac{1}{2}^{+})$&$\frac{1}{\sqrt{2}}b(cs-sc)$&$\chi_{MA}(23)$\\
$\Omega_{bc}^{'0}$&$0(\frac{1}{2}^{+})$&$\frac{1}{\sqrt{2}}b(cs+sc)$&$\chi_{MS}(23)$\\
$\Omega_{cc}^{*+}$&$0(\frac{3}{2}^{+})$&$ccs$&$\chi_{S}$\\
$\Omega_{c}^{*0}$&$0(\frac{3}{2}^{+})$&$css$&$\chi_{S}$\\
$\Omega_{bb}^{*-}$&$0(\frac{3}{2}^{+})$&$bbs$&$\chi_{S}$\\
$\Omega_{b}^{*-}$&$0(\frac{3}{2}^{+})$&$bss$&$\chi_{S}$\\
$\Omega_{bc}^{*0}$&$0(\frac{3}{2}^{+})$&$\frac{1}{\sqrt{2}}b(cs+sc)$&$\chi_{S}$\\
\bottomrule[1.00pt]
\bottomrule[1.00pt]
\end{tabular*}
\label{tab:Wavefunctions}
\end{table}

\begin{equation}
\begin{aligned}
\chi_{MS}(12)=\frac{1}{\sqrt{6}}(\uparrow\downarrow\uparrow+\downarrow\uparrow\uparrow-2\uparrow\uparrow\downarrow),
\end{aligned}
\label{eq:chiMS12}
\end{equation}
\begin{equation}
\begin{aligned}
\chi_{MS}(23)=\frac{1}{\sqrt{6}}(\uparrow\downarrow\uparrow+\uparrow\uparrow\downarrow-2\downarrow\uparrow\uparrow),
\end{aligned}
\label{eq:chiMS23}
\end{equation}
\begin{equation}
\begin{aligned}
\chi_{MA}(23)=\frac{1}{\sqrt{2}}(\uparrow\uparrow\downarrow-\uparrow\downarrow\uparrow).
\end{aligned}
\label{eq:chiMA23}
\end{equation}
Note that with our spin independent interaction, for the spin overlap between some baryons, we need the following formulae
\begin{equation}
\begin{aligned}
\left\langle\chi_{MS}(12)|\chi_{MS}(23)\right\rangle=-\frac{1}{2},
\end{aligned}
\label{eq:chiSchiS}
\end{equation}
\begin{equation}
\begin{aligned}
\left\langle\chi_{MS}(12)|\chi_{MA}(23)\right\rangle=-\frac{\sqrt{3}}{2}.
\end{aligned}
\label{eq:chiSchiA}
\end{equation}
Furthermore, for the $J^{P}=3/2^{+}$ ground baryon states in the special case $S_{z}=+3/2$, we obtain
\begin{equation}
\begin{aligned}
\chi_{S}=\uparrow\uparrow\uparrow.
\end{aligned}
\label{eq:chiS}
\end{equation}

\begin{figure}[htbp]
\centering
\includegraphics[width=0.8\linewidth,trim=150 580 250 120,clip]{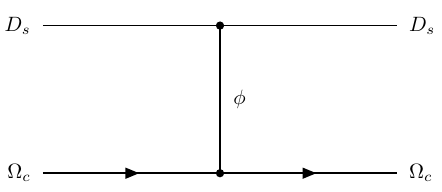} 
\caption{Diagram of the light vector meson $\phi$ exchange in the $D_{s}\Omega_{c}\rightarrow D_{s}\Omega_{c}$ transition.}
\label{fig:DsOmegac}
\end{figure} 

We can obtain the vertex $VBB$ with the wave functions of the baryons above.
Take the $D_{s}\Omega_{c}\rightarrow D_{s}\Omega_{c}$ process as an example, as shown in Fig. \ref{fig:DsOmegac}, where the interaction of the transition is contributed by the exchange of the light vector meson $\phi$ in the $t$-channel.
Thus, for the lower vertex we have
\begin{equation}
\begin{aligned}
\left\langle css \chi_{MS}(23)\left|gs\bar{s}\right|css \chi_{MS}(23)\right\rangle=g.
\end{aligned}
\label{eq:DsOmegac}
\end{equation}
One can see Refs. \cite{Sakai:2017avl,Dias:2018qhp} for more details. More details of the $P_{i}B_{i}\rightarrow P_{j}B_{j}$ and $V_{i}B_{i}\rightarrow V_{j}B_{j}$ transitions obtained through two vertices $VPP$ ($VVV$) and $VBB$ are given in the Appendix of Ref. \cite{Debastiani:2017ewu}, and we do not repeat them here. Note that we are concerned with the interaction near the threshold, so we have neglected the three-momentum of the external particles in the transitions. Therefore, for the polarizations of the two external vectors, we have $\epsilon_{1\mu}\epsilon_{3}^{\dagger\mu}=-\vec{\epsilon}_{1}\cdot\vec{\epsilon}_{3}^{\,\dagger}$ with $\epsilon^{0}=0$.

Finally, the explicit form of the potential from the vector meson exchange mechanism, as shown in Fig. \ref{fig:DsOmegac}, is given by
\begin{equation}
\begin{aligned}
v_{ij}=-C_{ij}\frac{1}{4f_{\pi}^{2}}(p_{i}^{0}+p_{j}^{0}),
\end{aligned}
\label{eq:vij1}
\end{equation}
where $p_{i}^{0}$ ($p_{j}^{0}$) is the energy of initial (final) meson. The coefficient matrix elements $C_{ij}=C_{ji}$, which are given in Tables \ref{tab:ccsss_1}-\ref{tab:bcsss_1}. Their relativistically corrected expression in the $s$-wave is obtained as \cite{Oset:2001cn}
\begin{equation}
\begin{aligned}
v_{ij}(\sqrt{s})=-C_{ij}\frac{2\sqrt{s}-M_{i}-M_{j}}{4f_{\pi}^{2}}\left(\frac{M_{i}+E_{i}}{2M_{i}}\right)^{1/2}\left(\frac{M_{j}+E_{j}}{2M_{j}}\right)^{1/2},
\end{aligned}
\label{eq:vij2}
\end{equation}

\noindent
where $M_{i}(M_{j})$ and $E_{i}(E_{j})$ denote the mass and center-of-mass energy of the initial (final) baryon, respectively.

\begin{table}[htbp]
\centering
\renewcommand\tabcolsep{0.1mm}
\renewcommand{\arraystretch}{1.50}
\caption{Coefficient matrix elements $C_{ij}$ in the $ccss\bar{s}$ sector.}
\begin{tabular*}{86mm}{@{\extracolsep{\fill}}l|ccc|ccc}
\toprule[1.00pt]
\toprule[1.00pt]
\multirow{3}{*}{$C_{ij}(PB(1/2^{+}),PB(3/2^{+}))$}&&$\eta^{'}\Omega_{cc}$&$D_{s}\Omega_{c}$&&$\eta^{'}\Omega_{cc}^{*}$&$D_{s}\Omega_{c}^{*}$\\
&$\eta^{'}\Omega_{cc}$&$0$&$\frac{\lambda}{\sqrt{6}}$&$\eta^{'}\Omega_{cc}^{*}$&$0$&$\frac{-\sqrt{2}\lambda}{\sqrt{3}}$\\
&$D_{s}\Omega_{c}$&$\frac{\lambda}{\sqrt{6}}$&$1$&$D_{s}\Omega_{c}^{*}$&$\frac{-\sqrt{2}\lambda}{\sqrt{3}}$&$1$\\
\hline
\multirow{3}{*}{$C_{ij}(VB(1/2^{+}),VB(3/2^{+}))$}&&$\phi\Omega_{cc}$&$D_{s}^{*}\Omega_{c}$&&$\phi\Omega_{cc}^{*}$&$D_{s}^{*}\Omega_{c}^{*}$\\
&$\phi\Omega_{cc}$&$0$&$\frac{\lambda}{2}$&$\phi\Omega_{cc}^{*}$&$0$&$-\lambda$\\
&$D_{s}^{*}\Omega_{c}$&$\frac{\lambda}{2}$&$1$&$D_{s}^{*}\Omega_{c}^{*}$&$-\lambda$&$1$\\
\bottomrule[1.00pt]
\bottomrule[1.00pt]
\end{tabular*}
\label{tab:ccsss_1}
\end{table}

\begin{table}[htbp]
\centering
\renewcommand\tabcolsep{0.1mm}
\renewcommand{\arraystretch}{1.50}
\caption{Coefficient matrix elements $C_{ij}$ in the $bbss\bar{s}$ sector.}
\begin{tabular*}{86mm}{@{\extracolsep{\fill}}l|ccc|ccc}
\toprule[1.00pt]
\toprule[1.00pt]
\multirow{3}{*}{$C_{ij}(PB(1/2^{+}),PB(3/2^{+}))$}&&$\eta^{'}\Omega_{bb}$&$\bar{B}_{s}\Omega_{b}$&&$\eta^{'}\Omega_{bb}^{*}$&$\bar{B}_{s}\Omega_{b}^{*}$\\
&$\eta^{'}\Omega_{bb}$&$0$&$0$&$\eta^{'}\Omega_{bb}^{*}$&$0$&$0$\\
&$\bar{B}_{s}\Omega_{b}$&$0$&$1$&$\bar{B}_{s}\Omega_{b}^{*}$&$0$&$1$\\
\hline
\multirow{3}{*}{$C_{ij}(VB(1/2^{+}),VB(3/2^{+}))$}&&$\phi\Omega_{bb}$&$\bar{B}_{s}^{*}\Omega_{b}$&&$\phi\Omega_{bb}^{*}$&$\bar{B}_{s}^{*}\Omega_{b}^{*}$\\
&$\phi\Omega_{bb}$&$0$&$0$&$\phi\Omega_{bb}^{*}$&$0$&$0$\\
&$\bar{B}_{s}^{*}\Omega_{b}$&$0$&$1$&$\bar{B}_{s}^{*}\Omega_{b}^{*}$&$0$&$1$\\
\bottomrule[1.00pt]
\bottomrule[1.00pt]
\end{tabular*}
\label{tab:bbsss_1}
\end{table}

\begin{table}[htbp]
\centering
\renewcommand\tabcolsep{2mm}
\renewcommand{\arraystretch}{1.50}
\caption{Coefficient matrix elements $C_{ij}$ in the $bcss\bar{s}$ sector.}
\begin{tabular*}{86mm}{@{\extracolsep{\fill}}l|ccccc}
\toprule[1.00pt]
\toprule[1.00pt]
\multirow{5}{*}{$C_{ij}(PB(1/2^{+}))$}&&$\eta^{'}\Omega_{bc}$&$\eta^{'}\Omega_{bc}^{'}$&$D_{s}\Omega_{b}$&$\bar{B}_{s}\Omega_{c}$\\
&$\eta^{'}\Omega_{bc}$&$0$&$0$&$0$&$0$\\
&$\eta^{'}\Omega_{bc}^{'}$&$0$&$0$&$\frac{-2\lambda}{\sqrt{3}}$&$0$\\
&$D_{s}\Omega_{b}$&$0$&$\frac{-2\lambda}{\sqrt{3}}$&$1$&$0$\\
&$\bar{B}_{s}\Omega_{c}$&$0$&$0$&$0$&$1$\\
\hline
\multirow{4}{*}{$C_{ij}(PB(3/2^{+}))$}&&$\eta^{'}\Omega_{bc}^{*}$&$D_{s}\Omega_{b}^{*}$&$\bar{B}_{s}\Omega_{c}^{*}$&\\
&$\eta^{'}\Omega_{bc}^{*}$&$0$&$\frac{-2\lambda}{\sqrt{3}}$&$0$&\\
&$D_{s}\Omega_{b}^{*}$&$\frac{-2\lambda}{\sqrt{3}}$&$1$&$0$&\\
&$\bar{B}_{s}\Omega_{c}^{*}$&$0$&$0$&$1$&\\
\hline
\multirow{5}{*}{$C_{ij}(VB(1/2^{+}))$}&&$\phi\Omega_{bc}$&$\phi\Omega_{bc}^{'}$&$D_{s}^{*}\Omega_{b}$&$\bar{B}_{s}^{*}\Omega_{c}$ \\
&$\phi\Omega_{bc}$&$0$&$0$&$0$&$0$\\
&$\phi\Omega_{bc}^{'}$&$0$&$0$&$-\sqrt{2}\lambda$&$0$\\
&$D_{s}^{*}\Omega_{b}$&$0$&$-\sqrt{2}\lambda$&$1$&$0$\\
&$\bar{B}_{s}^{*}\Omega_{c}$&$0$&$0$&$0$&$1$\\
\hline
\multirow{5}{*}{$C_{ij}(VB(3/2^{+}))$}&&$\phi\Omega_{bc}^{*}$&$D_{s}^{*}\Omega_{b}^{*}$&$\bar{B}_{s}^{*}\Omega_{c}^{*}$&\\
&$\phi\Omega_{bc}^{*}$&$0$&$-\sqrt{2}\lambda$&$0$&\\
&$D_{s}^{*}\Omega_{b}^{*}$&$-\sqrt{2}\lambda$&$1$&$0$&\\
&$\bar{B}_{s}^{*}\Omega_{c}^{*}$&$0$&$0$&$1$&\\
\bottomrule[1.00pt]
\bottomrule[1.00pt]
\end{tabular*}
\label{tab:bcsss_1}
\end{table}

In Tables \ref{tab:ccsss_1}-\ref{tab:bcsss_1}, we keep the contributions of the light vector meson and $D_{s}^{*}$ meson exchange, ignoring the heavier ones $J/\psi$, $B_{s}^{*}$, etc. Compared to the light vector mesons, the contribution of exchanging $D_{s}^{*}$ meson has a suppression factor of the order of $m_{V}^{2}/m_{D_{s}^{*}}^{2}$, i.e., $\lambda$, which is approximated as $0.25$ in the numerical calculations, see more details in Ref. \cite{Debastiani:2017ewu}.

With the potentials of the $s$-wave, one can obtain the scattering amplitudes by solving the Bethe-Salpeter equation with the on-shell approximation \cite{Oset:1997it}, given by
\begin{equation}
\begin{aligned} 
T = [1-vG]^{-1}v.
\end{aligned}
\label{eq:BSE}
\end{equation}
The elements of the diagonal matrix $G$ in Eq. \eqref{eq:BSE} are the loop functions of intermediate mesons and baryons, whose the expression is given by
\begin{equation}
\begin{aligned} 
G_{l}=i \int \frac{d^{4} q}{(2 \pi)^{4}}\frac{2M_{l}}{(P-q)^{2}-M_{l}^{2}+i \epsilon} \frac{1}{q^{2}-m_{l}^{2}+i \epsilon},
\end{aligned}
\label{eq:G}
\end{equation}
where $P=k_{1}+p_{1}$ is the total momentum of the meson and baryon coupled system. The loop function of Eq. \eqref{eq:G} is logarithmically divergent, and therefore it should be regularized, {\it i.e.}, using the regularization of the three-momentum cutoff method \cite{Oset:1997it}, one obtains
\begin{equation}
\begin{aligned}
G_{l}(s)=\int_{0}^{q_{max}} \frac{q^{2} d q}{2 \pi^{2}} \frac{1}{2 \omega_{l}(q)} \frac{M_{l}}{E_{l}(q)} \frac{1}{p^{0}+k^{0}-\omega_{l}(q)-E_{l}(q)+i \epsilon},
\end{aligned}
\label{eq:GCO}
\end{equation}
where we define $p^{0}+k^{0}=\sqrt{s}$, $\omega_{l}(q)=\sqrt{q^2+m_{l}^2}$, and $E_{l}(q)=\sqrt{q^2+M_{l}^2}$, with $m_{l}$ and $M_{l}$ the masses of meson and baryon of the $l$ channel, respectively, and $q_{max}$ the free parameter. Furthermore, the divergence can also be treated with the dimensional regularization scheme \cite{Oller:2000fj,Jido:2003cb}, having
\begin{equation}
\begin{aligned}
G_{l}(s)=& \frac{2 M_{l}}{16 \pi^{2}}\left\{a(\mu)+\ln \frac{M_{l}^{2}}{\mu^{2}}+\frac{m_{l}^{2}-M_{l}^{2}+s}{2 s} \ln \frac{m_{l}^{2}}{M_{l}^{2}}\right.\\
&+\frac{q_{cml}(s)}{\sqrt{s}}\left[\ln \left(s-\left(M_{l}^{2}-m_{l}^{2}\right)+2 q_{cml}(s) \sqrt{s}\right)\right.\\
&+\ln \left(s+\left(M_{l}^{2}-m_{l}^{2}\right)+2 q_{cml}(s) \sqrt{s}\right) \\
&-\ln \left(-s-\left(M_{l}^{2}-m_{l}^{2}\right)+2 q_{cml}(s) \sqrt{s}\right) \\
&\left.\left.-\ln \left(-s+\left(M_{l}^{2}-m_{l}^{2}\right)+2 q_{cml}(s) \sqrt{s}\right)\right]\right\}.
\end{aligned}
\label{eq:GDR}
\end{equation}
Here, $\mu$ is the regularization scale, $a(\mu)$ the subtraction constant, and $q_{cml}(s)$ the three momentum of the particle in the center-of-mass frame, given by 
\begin{equation}
\begin{aligned}
q_{cml}(s)=\frac{\lambda^{1 / 2}\left(s, M_{l}^{2}, m_{l}^{2}\right)}{2 \sqrt{s}}
\end{aligned}
\end{equation}
with the usual K\"all\'en triangle function $\lambda(a, b, c)=a^{2}+b^{2}+c^{2}-2(a b+a c+b c)$.
In the present work, we will use the dimensional regularization scheme. The parameter of $\mu$ is chosen as $\mu=q_{max}=650$ MeV, which is given in Ref. \cite{Debastiani:2017ewu} by reproducing the three $\Omega_{c}$ states found in Ref. \cite{LHCb:2017uwr}. And $a(\mu)$ is determined by matching the values of $G_{l}(s)$ function from these two methods at the threshold, as in Ref. \cite{Oset:2001cn}.

In the ChUA, to search for the pole, one must extrapolate the $G_{l}(s)$ function into the second Riemann sheet for $\text{Re}(\sqrt{s})$ being greater than the threshold of the $l$ channel ($m_{th}$), given by
\begin{equation}
\begin{aligned}
G_{l}^{(II)}(s)&=G_{l}(s)-2i \text{Im}G_{l}(s) \\
&=G_{l}(s)+\frac{i}{2\pi}\frac{M_{l}q_{cml}(s)}{\sqrt{s}}.
\end{aligned}
\end{equation}
Thus, $G_{l}^{(II)}(s)$ is taken when $\text{Re}(\sqrt{s}) \geq m_{th}$, where the corresponding channel is open for the decay channel, see the results later.
Note that the poles on the real energy axis and below the threshold on the first Riemann sheet
are bound states. In the coupled channels approach, the quasi-bound states appear as poles off the real energy axis on the unphysical Riemann sheet. 

In addition, the coupling constants $g_{i}$ can be calculated from the Laurent expansion of the amplitudes near the pole $\sqrt{s_{p}}$ for different channels \cite{Yamagata-Sekihara:2010kpd}, written as
\begin{equation}
T_{ij}=\frac{g_{i}g_{j}}{\sqrt{s}-\sqrt{s_{p}}}.
\end{equation}

\section{Numerical results}\label{sec:Results}

Table \ref{tab:Thresholds} shows the thresholds of different channels. The meson and baryon masses used in the calculations are taken from Refs. \cite{Zhou:2018bkn,ParticleDataGroup:2022pth,Yu:2022ymb}. In Tables \ref{tab:ccsss}-\ref{tab:bcsss}, we list the pole positions and their coupling constants for each channel. In brackets after the values of the pole position, the $+$ sign indicates that the corresponding channel is closed, and the $-$ sign indicates that the corresponding channel is open. The largest coupling constant is highlighted in bold. As mentioned in the previous section, we take the free parameter $\mu=650$ MeV for all systems.

\begin{table*}[htbp]
\centering
\renewcommand\tabcolsep{2mm}
\renewcommand{\arraystretch}{1.50}
\caption{Thresholds (in MeV) of the relevant channels.}
\begin{tabular*}{178mm}{@{\extracolsep{\fill}}lcccccccc}
\toprule[1.00pt]
\toprule[1.00pt]
\multicolumn{9}{c}{$ccss\bar{s}$}\\
\hline
Channels&$\eta^{'}\Omega_{cc}$&$D_{s}\Omega_{c}$&$\eta^{'}\Omega_{cc}^{*}$&$D_{s}\Omega_{c}^{*}$&$\phi\Omega_{cc}$&$D_{s}^{*}\Omega_{c}$&$\phi\Omega_{cc}^{*}$&$D_{s}^{*}\Omega_{c}^{*}$\\ 
Thresholds&$4672.78$&$4663.55$&$4729.78$&$4734.25$&$4734.46$&$4807.40$&$4791.46$&$4878.10$\\
\hline
\multicolumn{9}{c}{$bbss\bar{s}$}\\
\hline
Channels&$\eta^{'}\Omega_{bb}$&$\bar{B}_{s}\Omega_{b}$&$\eta^{'}\Omega_{bb}^{*}$&$\bar{B}_{s}\Omega_{b}^{*}$&$\phi\Omega_{bb}$&$\bar{B}_{s}^{*}\Omega_{b}$&$\phi\Omega_{bb}^{*}$&$\bar{B}_{s}^{*}\Omega_{b}^{*}$\\ 
Thresholds&$11187.78$&$11412.12$&$11215.78$&$11435.92$&$11249.46$&$11460.60$&$11277.46$&$11484.40$\\ 
\hline
\multicolumn{9}{c}{$bcss\bar{s}$}\\
\hline
Channels&$\eta^{'}\Omega_{bc}$&$\eta^{'}\Omega_{bc}^{'}$&$D_{s}\Omega_{b}$&$\bar{B}_{s}\Omega_{c}$&$\eta^{'}\Omega_{bc}^{*}$&$D_{s}\Omega_{b}^{*}$&$\bar{B}_{s}\Omega_{c}^{*}$&\\ 
Thresholds&$7968.78$&$8004.78$&$8013.55$&$8062.12$&$8023.78$&$8037.35$&$8132.82$&\\ 
Channels&$\phi\Omega_{bc}$&$\phi\Omega_{bc}^{'}$&$D_{s}^{*}\Omega_{b}$&$\bar{B}_{s}^{*}\Omega_{c}$&$\phi\Omega_{bc}^{*}$&$D_{s}^{*}\Omega_{b}^{*}$&$\bar{B}_{s}^{*}\Omega_{c}^{*}$&\\ 
Thresholds&$8030.46$&$8066.46$&$8157.40$&$8110.60$&$8085.46$&$8181.20$&$8181.30$&\\
\bottomrule[1.00pt]
\bottomrule[1.00pt]
\end{tabular*}
\label{tab:Thresholds}
\end{table*}

\begin{figure}[htbp]
\begin{minipage}{0.49\linewidth}
\centering
\includegraphics[width=1\linewidth,trim=0 0 0 0,clip]{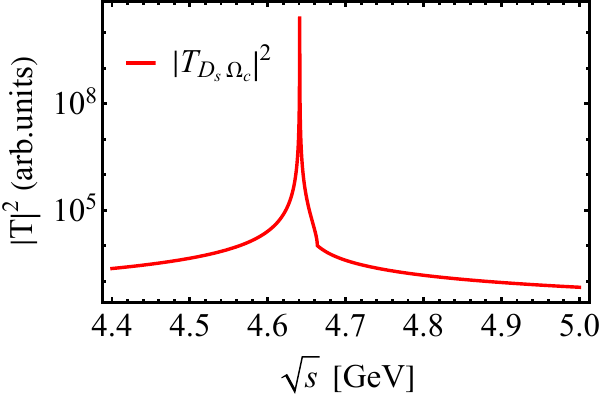} 
\label{fig:ccsss_1}
\end{minipage}
\begin{minipage}{0.49\linewidth} 
\centering 
\includegraphics[width=1\linewidth,trim=0 0 0 0,clip]{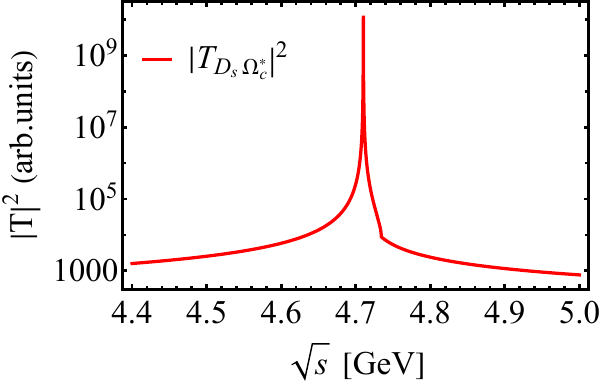} 
\label{fig:ccsss_2}  
\end{minipage}	
\begin{minipage}{0.49\linewidth} 
\centering 
\includegraphics[width=1\linewidth,trim=0 0 0 0,clip]{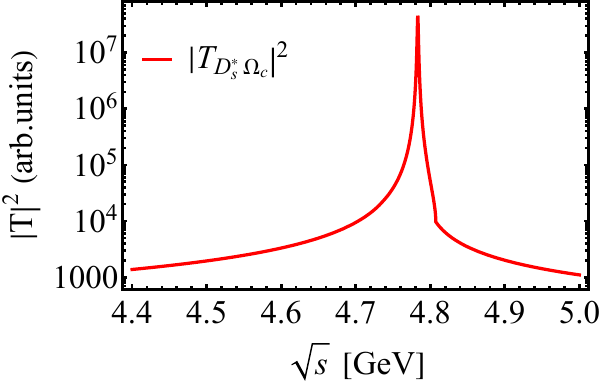} 
\label{fig:ccsss_3}  
\end{minipage}
\begin{minipage}{0.49\linewidth} 
\centering 
\includegraphics[width=1\linewidth,trim=0 0 0 0,clip]{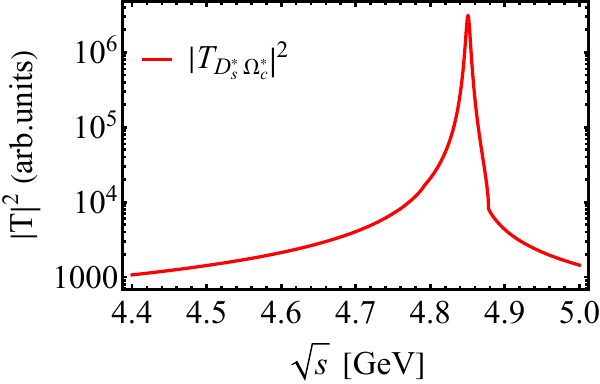} 
\label{fig:ccsss_4}  
\end{minipage}
\caption{Modulus square of the amplitudes in the $ccss\bar{s}$ sector, where the ordinate is a logarithmic coordinate system.}
\label{fig:ccsss}
\end{figure}

\begin{table}[htbp]
\centering
\renewcommand\tabcolsep{0.8mm}
\renewcommand{\arraystretch}{1.50}
\caption{Poles (in MeV) and their couplings for each channel in the $ccss\bar{s}$ sector. The sign $+$ indicates the corresponding channel is closed, and the sign $-$ indicates that the corresponding channel is open, in the same order as the channels in Table \ref{tab:coupledchannels}. The largest coupling constant is highlighted in bold.}
\begin{tabular*}{86mm}{@{\extracolsep{\fill}}lccc}
\toprule[1.00pt]
\toprule[1.00pt]
$I(J^{P})$&Poles position&\multicolumn{2}{c}{Couplings}\\
\hline
$0(\frac{1}{2}^{-})$&$4640.95$ $(++)$&$|g_{\eta^{'}\Omega_{cc}}|=0.16$&$|g_{D_{s}\Omega_{c}}|=\bf{2.18}$\\  
$0(\frac{3}{2}^{-})$&$4710.21$ $(++)$&$|g_{\eta^{'}\Omega_{cc}^{*}}|=0.32$&$|g_{D_{s}\Omega_{c}^{*}}|=\bf{2.17}$\\  
$0(\frac{1}{2}^{-},\frac{3}{2}^{-})$&$4783.23-0.74i$ $(-+)$&$|g_{\phi\Omega_{cc}}|=0.21$&$|g_{D_{s}^{*}\Omega_{c}}|=\bf{2.24}$\\ 
$0(\frac{1}{2}^{-},\frac{3}{2}^{-},\frac{5}{2}^{-})$&$4850.69-3.01i$ $(-+)$&$|g_{\phi\Omega_{cc}^{*}}|=0.39$&$|g_{D_{s}^{*}\Omega_{c}^{*}}|=\bf{2.30}$\\
\bottomrule[1.00pt]
\bottomrule[1.00pt]
\end{tabular*}
\label{tab:ccsss}
\end{table}

First, in Fig. \ref{fig:ccsss}, we have plotted the modulus square of the amplitudes $|T|^{2}$ of the diagonal matrix elements in the $ccss\bar{s}$ system, where four peak structures are clearly visible. Correspondingly, we find four poles on the complex Riemann sheet, which are listed in Table \ref{tab:ccsss}. The first two poles $4640.95$ MeV and $4710.21$ MeV are located on the first Riemann sheet without imaginary parts, indicating that the widths of these two states are zero and that these two states cannot decay into their coupled channel. They are mainly coupled to the $D_{s}\Omega_{c}$ channel with quantum number $I(J^{P})=0(1/2^{-})$ and  the $D_{s}\Omega_{c}^{*}$ channel with $I(J^{P})=0(3/2^{-})$, respectively, where the first state could mostly qualify as a $D_{s}\Omega_{c}$ molecule and the second one as a $D_{s}\Omega_{c}^{*}$ molecule, and both of them with a binding of about $23-24$ MeV. The latter two poles are around $24-27$ MeV below the thresholds of the most relevant channels, which are found on the second Riemann sheet and the bound states of the $D_{s}^{*}\Omega_{c}$ and $D_{s}^{*}\Omega_{c}^{*}$ channels, respectively, with the widths of twice the imaginary part of the pole.
These two resonances can decay separately into the open channels $\phi\Omega_{cc}$ and $\phi\Omega_{cc}^{*}$. 

Note that Ref. \cite{Hofmann:2005sw} found a state located at $4571$ MeV with a width of $90$ MeV and mainly coupled to the $D_{s}\Omega_{c}$ channel based on the interactions from the chiral effective Lagrangians. 
As commented in Ref. \cite{Marse-Valera:2022khy}, the subtraction scales $\mu$ used in Ref. \cite{Hofmann:2005sw} is unreasonably large to ensure that the heavy channel has a negligible effect on the low-energy scattering of the light channel.
In Ref. \cite{Dong:2021bvy} four states with binding energies around $0.1$ MeV to $15$ MeV were found with single channel interactions based on the interactions from the heavy quark spin symmetry.
The binding energies of the bound states obtained by us are slightly higher than theirs.
A main reason for this is that we use the regularization scale $\mu=650$ MeV in the model, which is obtained by fitting the experimental data in Ref. \cite{Debastiani:2017ewu}.
In addition, the one-boson-exchange model was used to study the mass spectra of the $D_{s}^{(*)}\Omega_{c}^{(*)}$-type molecular states in Ref. \cite{Wang:2023aob}, where they obtained similar results using a slightly larger cutoff $\Lambda$ in the calculations.

\begin{figure}[htbp]
\begin{minipage}{0.49\linewidth}
\centering
\includegraphics[width=1\linewidth,trim=0 0 0 0,clip]{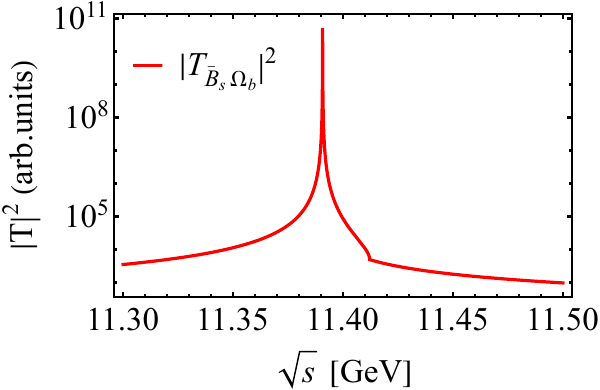} 
\label{fig:bbsss_1}
\end{minipage}
\begin{minipage}{0.49\linewidth} 
\centering 
\includegraphics[width=1\linewidth,trim=0 0 0 0,clip]{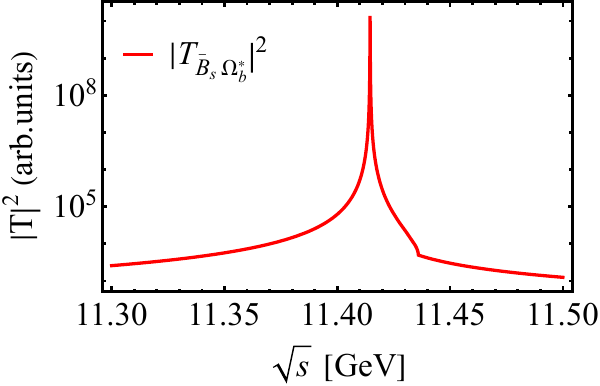} 
\label{fig:bbsss_2}  
\end{minipage}	
\begin{minipage}{0.49\linewidth} 
\centering 
\includegraphics[width=1\linewidth,trim=0 0 0 0,clip]{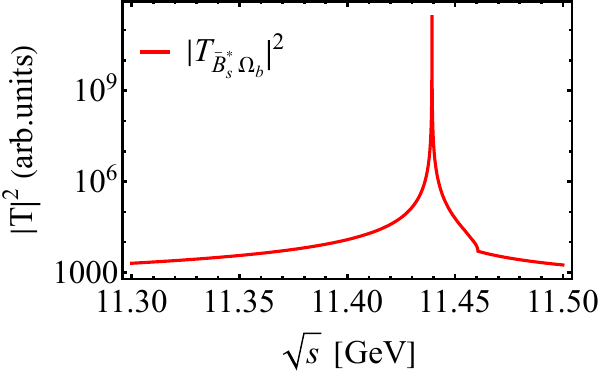} 
\label{fig:bbsss_3}  
\end{minipage}
\begin{minipage}{0.49\linewidth} 
\centering 
\includegraphics[width=1\linewidth,trim=0 0 0 0,clip]{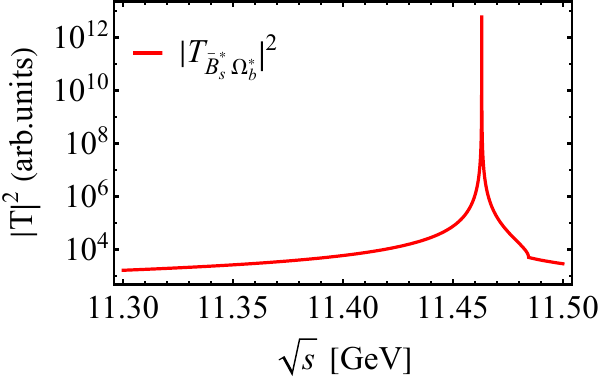} 
\label{fig:bbsss_4}  
\end{minipage}
\caption{Modulus square of amplitudes in the $bbss\bar{s}$ sector.}
\label{fig:bbsss}
\end{figure}

\begin{table}[htbp]
\centering
\renewcommand\tabcolsep{0.8mm}
\renewcommand{\arraystretch}{1.50}
\caption{Poles (in MeV) and their couplings for every channel in the $bbss\bar{s}$ sector.}
\begin{tabular*}{86mm}{@{\extracolsep{\fill}}lccc}
\toprule[1.00pt]
\toprule[1.00pt]
$I(J^{P})$&Poles position&\multicolumn{2}{c}{Couplings}\\
\hline
$0(\frac{1}{2}^{-})$&$11390.73$ $(++)$&$|g_{\eta^{'}\Omega_{bb}}|=0.00$&$|g_{\bar{B}_{s}\Omega_{b}}|=\bf{1.77}$\\  
$0(\frac{3}{2}^{-})$&$11414.56$ $(++)$&$|g_{\eta^{'}\Omega_{bb}^{*}}|=0.00$&$|g_{\bar{B}_{s}\Omega_{b}^{*}}|=\bf{1.77}$ \\  
$0(\frac{1}{2}^{-},\frac{3}{2}^{-})$&$11439.27$ $(++)$&$|g_{\phi\Omega_{bb}}|=0.00$&$|g_{\bar{B}_{s}^{*}\Omega_{b}}|=\bf{1.77}$\\  
$0(\frac{1}{2}^{-},\frac{3}{2}^{-},\frac{5}{2}^{-})$&$11463.09$ $(++)$&$|g_{\phi\Omega_{bb}^{*}}|=0.00$&$|g_{\bar{B}_{s}^{*}\Omega_{b}^{*}}|=\bf{1.77}$\\
\bottomrule[1.00pt]
\bottomrule[1.00pt]
\end{tabular*}
\label{tab:bbsss}
\end{table}

\begin{figure*}[htbp]
\begin{minipage}{0.24\linewidth}
\centering
\includegraphics[width=1\linewidth,trim=0 0 0 0,clip]{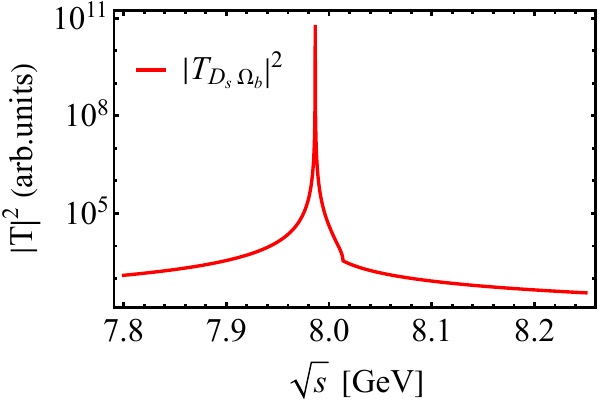} 
\label{fig:bcsss_1}
\end{minipage}
\begin{minipage}{0.24\linewidth} 
\centering 
\includegraphics[width=1\linewidth,trim=0 0 0 0,clip]{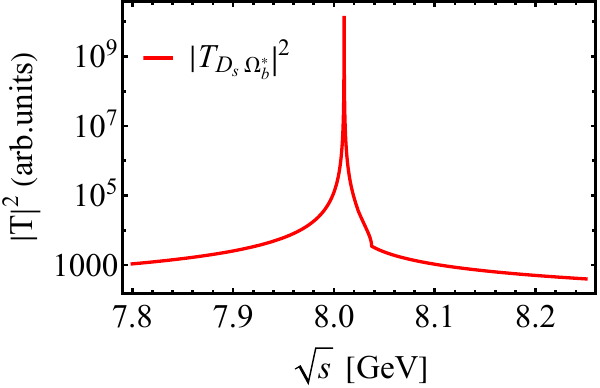} 
\label{fig:bcsss_2}  
\end{minipage}	
\begin{minipage}{0.24\linewidth} 
\centering 
\includegraphics[width=1\linewidth,trim=0 0 0 0,clip]{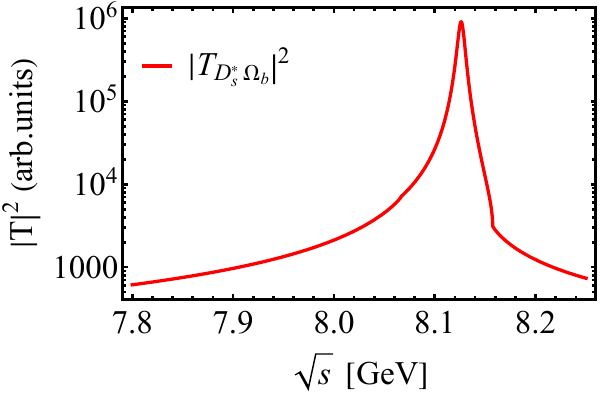} 
\label{fig:bcsss_3}  
\end{minipage}
\begin{minipage}{0.24\linewidth} 
\centering 
\includegraphics[width=1\linewidth,trim=0 0 0 0,clip]{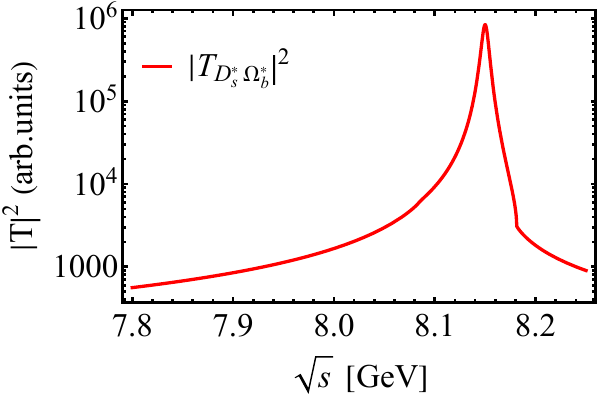} 
\label{fig:bcsss_4}  
\end{minipage}
\begin{minipage}{0.24\linewidth} 
\centering 
\includegraphics[width=1\linewidth,trim=0 0 0 0,clip]{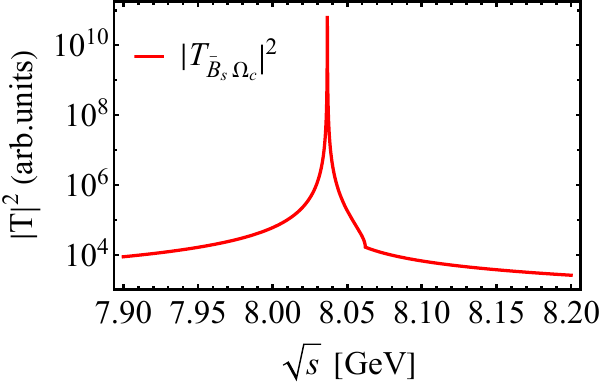} 
\label{fig:bcsss_5}  
\end{minipage}
\begin{minipage}{0.24\linewidth} 
\centering 
\includegraphics[width=1\linewidth,trim=0 0 0 0,clip]{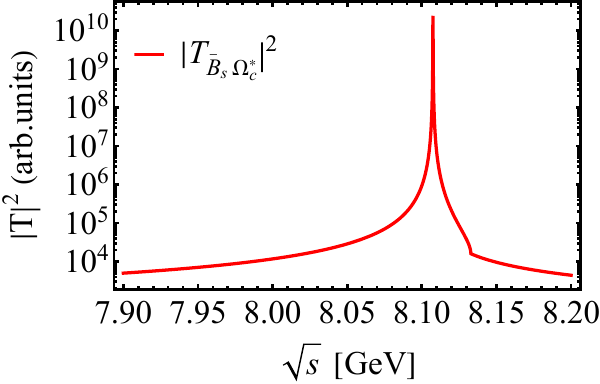} 
\label{fig:bcsss_6}  
\end{minipage}
\begin{minipage}{0.24\linewidth} 
\centering 
\includegraphics[width=1\linewidth,trim=0 0 0 0,clip]{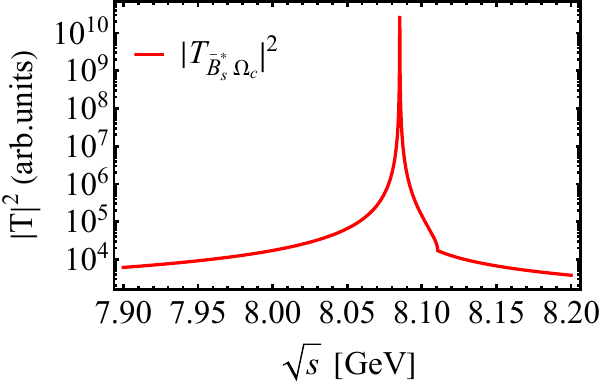} 
\label{fig:bcsss_7}  
\end{minipage}
\begin{minipage}{0.24\linewidth} 
\centering 
\includegraphics[width=1\linewidth,trim=0 0 0 0,clip]{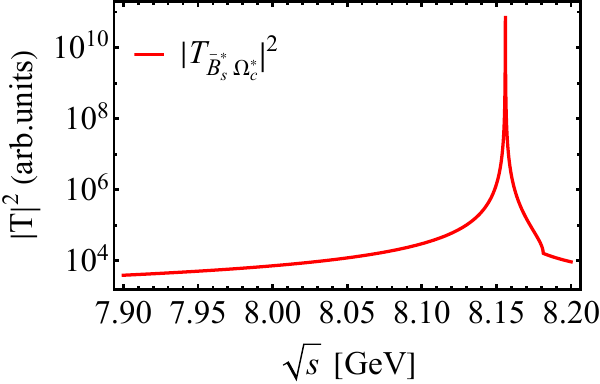} 
\label{fig:bcsss_8}  
\end{minipage}
\caption{Modulus square of amplitudes in the $bcss\bar{s}$ sector.}
\label{fig:bcsss}
\end{figure*}	

\begin{table*}[htbp]
\centering
\renewcommand\tabcolsep{2mm}
\renewcommand{\arraystretch}{1.50}
\caption{Poles (in MeV) and their couplings for every channel in the $bcss\bar{s}$ sector.}
\begin{tabular*}{178mm}{@{\extracolsep{\fill}}lccccc}
\toprule[1.00pt]
\toprule[1.00pt]
$I(J^{P})$&Poles position&\multicolumn{4}{c}{Couplings}\\
\hline
\multirow{2}{*}{$0(\frac{1}{2}^{-})$}&$7986.54$ $(++++)$&$|g_{\eta^{'}\Omega_{bc}}|=0.00$&$|g_{\eta^{'}\Omega_{bc}^{'}}|=0.36$&$|g_{D_{s}\Omega_{b}}|=\bf{1.81}$&$|g_{\bar{B}_{s}\Omega_{c}}|=0.00$\\  
&$8036.70$ $(++++)$&$|g_{\eta^{'}\Omega_{bc}}|=0.00$&$|g_{\eta^{'}\Omega_{bc}^{'}}|=0.00$&$|g_{D_{s}\Omega_{b}}|=0.00$&$|g_{\bar{B}_{s}\Omega_{c}}|=\bf{2.63}$\\ 
\hline
\multirow{2}{*}{$0(\frac{3}{2}^{-})$}&$8010.12$ $(+++)$&$|g_{\eta^{'}\Omega_{bc}^{*}}|=0.36$&$|g_{D_{s}\Omega_{b}^{*}}|=\bf{1.80}$&$|g_{\bar{B}_{s}\Omega_{c}^{*}}|=0.00$&\\ 
&$8107.45$ $(+++)$&$|g_{\eta^{'}\Omega_{bc}^{*}}|=0.00$&$|g_{D_{s}\Omega_{b}^{*}}|=0.00$&$|g_{\bar{B}_{s}\Omega_{c}^{*}}|=\bf{2.59}$&\\ 
\hline
\multirow{2}{*}{$0(\frac{1}{2}^{-},\frac{3}{2}^{-})$}&$8126.30-3.82i$ $(--++)$&$|g_{\phi\Omega_{bc}}|=0.00$&$|g_{\phi\Omega_{bc}^{'}}|=0.41$&$|g_{D_{s}^{*}\Omega_{b}}|=\bf{1.91}$&$|g_{\bar{B}_{s}^{*}\Omega_{c}}|=0.00$\\  
&$8085.20$ $(++++)$&$|g_{\phi\Omega_{bc}}|=0.00$&$|g_{\phi\Omega_{bc}^{'}}|=0.00$&$|g_{D_{s}^{*}\Omega_{b}}|=0.00$&$|g_{\bar{B}_{s}^{*}\Omega_{c}}|=\bf{2.63}$\\
\hline
\multirow{2}{*}{$0(\frac{1}{2}^{-},\frac{3}{2}^{-},\frac{5}{2}^{-})$}&$8150.07-3.97i$ $(-++)$&$|g_{\phi\Omega_{bc}^{*}}|=0.41$&$|g_{D_{s}^{*}\Omega_{b}^{*}}|=\bf{1.91}$&$|g_{\bar{B}_{s}^{*}\Omega_{c}^{*}}|=0.00$&\\ 
&$8155.95$ $(+++)$&$|g_{\phi\Omega_{bc}^{*}}|=0.00$&$|g_{D_{s}^{*}\Omega_{b}^{*}}|=0.00$&$|g_{\bar{B}_{s}^{*}\Omega_{c}^{*}}|=\bf{2.60}$&\\   
\bottomrule[1.00pt]
\bottomrule[1.00pt]
\end{tabular*}
\label{tab:bcsss}
\end{table*}

Next, in the $bbss\bar{s}$ sector, we show our results in Fig. \ref{fig:bbsss}. From the modulus square of the amplitudes we can see four distinct peaks around the energy range $11380$ MeV to $11480$ MeV. The poles and their coupling constants for each coupled channel are given in Table \ref{tab:bbsss}. The four states obtained are the $\bar{B}_{s}\Omega_{b}$, $\bar{B}_{s}\Omega_{b}^{*}$, $\bar{B}_{s}^{*}\Omega_{b}$, and $\bar{B}_{s}^{*}\Omega_{b}^{*}$ bound states, respectively, which are all found on the first Riemann sheet. Due to the absence of the exchange of light pseudoscalar and $B_{s}^{*}$ mesons, which means that there are no coupled channels for each other, see the coefficients in Table \ref{tab:bbsss_1}, the obtained poles are all without imaginary parts and all the couplings of the lower channels are zero. This is different from the case of the $ccss\bar{s}$ sector with the $D_{s}^{*}$ exchange, see the results in Table \ref{tab:ccsss}. The thresholds of the $\bar{B}_{s}\Omega_{b}$, $\bar{B}_{s}\Omega_{b}^{*}$, $\bar{B}_{s}^{*}\Omega_{b}$, and $\bar{B}_{s}^{*}\Omega_{b}^{*}$ channels are $11412.12$ MeV, $11435.92$ MeV, $11460.60$ MeV, and $11484.40$ MeV, respectively, see Table \ref{tab:Thresholds}, and thus we have these states all bound at about $21$ MeV, because of in fact the similar single channel interaction.

Finally, there are eight bound states in the $bcss\bar{s}$ sector, as shown in Fig. \ref{fig:bcsss} and Table \ref{tab:bcsss}. In the case of $PB(1/2^{+})$, two poles couple most strongly to the channels $D_{s}\Omega_{b}$ and $\bar{B}_{s}\Omega_{c}$, with the binding energies $27$ MeV and $25$ MeV, respectively. In the case of $PB(3/2^{+})$, the pole at $8010.12$ MeV would be mostly a $D_{s}\Omega_{b}^{*}$ bound state with a binding energy of $27$ MeV, and the pole at $8107.45$ MeV would be mostly a $\bar{B}_{s}\Omega_{c}^{*}$ state bound at $25$ MeV. Note that these states have no width, although for the last one, $8107.45$ MeV, there may be two lower channels open that are not coupled to the $\bar{B}_{s}\Omega_{c}^{*}$ channel. The two states obtained in the case of $VB(1/2^{+})$ sector degenerate in $J^{P}=1/2^{-}$ and $J^{P}=3/2^{-}$. The one of $8126.30-3.82i$ MeV is a molecule of $D_{s}^{*}\Omega_{b}$ and can decay into the open channel $\phi\Omega_{bc}^{'}$. The other one of $8085.20$ MeV is a molecule of $\bar{B}_{s}^{*}\Omega_{c}$ with zero width, which is not coupled to the other channels in this sector. Furthermore, the two states obtained in the case of the $VB(3/2^{+})$ sector degenerate in $J^{P}=1/2^{-}$, $3/2^{-}$, and $5/2^{-}$. The state $8150.07-3.97i$ MeV is located below the threshold of the most relevant channel $D_{s}^{*}\Omega_{b}^{*}$ with the largest coupling constant $|g_{D_{s}^{*}\Omega_{b}^{*}}|=1.91$, where its binding energy is $31$ MeV, and the decay channel is $\phi\Omega_{bc}^{*}$. The last state in Table \ref{tab:bcsss} can be considered as a $\bar{B}_{s}^{*}\Omega_{c}^{*}$ bound state with a binding energy of $25$ MeV and the coupling $|g_{\bar{B}_{s}^{*}\Omega_{c}^{*}}|=2.60$, which has also no width since there are no coupled channels to decay to.

Note that for the molecular states obtained above, the attractions of their most relevant channels are provided by the exchange of the vector meson $\phi$, while the interactions in nondiagonal terms are weaker due to the exchange of the heavy vector meson $D_{s}^{*}$.
It is very interesting that in the absence of long-range interactions mediated by the exchange of $\pi$ meson, the bound states can only be formed by the exchange of the heavier meson $\phi$. 
Thus, the binding energies of these systems are only tens MeV, which become more reasonable in the molecular picture for the heavy sector. It should be mentioned that in some cases, the deep bound systems are also found, such as the $\Lambda_c (2595)$ state in the $D N$ system~\cite{Hofmann:2005sw,Mizutani:2006vq} and the $f_2(1270)$ in the $\rho \rho$ interaction~\cite{Molina:2008jw}.
Therefore, these types of doubly heavy pentaquarks with the most strange quark contents $ccss\bar{s}$, $bbss\bar{s}$, and $bcss\bar{s}$ are excellent systems to validate this mechanism.
We look forward to the experimental search for these states to confirm the model used in this paper.
If these states exist, it will deepen our understanding of the nonperturbative behavior of the strong interaction.

\section{Summary}\label{sec:Conclusions}

In the present work, we make a study of $\Omega_{cc}$, $\Omega_{bb}$, and $\Omega_{bc}$ molecular states dynamically generated from the meson-baryon interaction in the charm and bottom sectors. 
With the extended local hidden gauge approach, we obtain the dominant interactions through the vector meson exchange mechanism, and then calculate the scattering amplitudes by solving the Bethe-Salpeter equation in the on-shell approximation. 
By searching for poles in the complex energy plane, the masses and widths of the molecular states are determined.
We find several candidates for hadronic molecular states with the quark components $ccss\bar{s}$, $bbss\bar{s}$, and $bcss\bar{s}$.

In the $ccss\bar{s}$ sector, we obtain four bound states mainly coupled to the $D_{s}\Omega_{c}$, $D_{s}\Omega_{c}^{*}$, $D_{s}^{*}\Omega_{c}$, and $D_{s}^{*}\Omega_{c}^{*}$ channels, respectively.
In the $bbss\bar{s}$ sector, we obtain four bound states mainly coupled to the $\bar{B}_{s}\Omega_{b}$, $\bar{B}_{s}\Omega_{b}^{*}$, $\bar{B}_{s}^{*}\Omega_{b}$, and $\bar{B}_{s}^{*}\Omega_{b}^{*}$ channels, respectively.
In the $bcss\bar{s}$ sector, we obtain eight bound states mainly coupled to $D_{s}\Omega_{b}$, $\bar{B}_{s}\Omega_{c}$, $D_{s}\Omega_{b}^{*}$, $\bar{B}_{s}\Omega_{c}^{*}$, $D_{s}^{*}\Omega_{b}$, $\bar{B}_{s}^{*}\Omega_{c}$, $D_{s}^{*}\Omega_{b}^{*}$, and $\bar{B}_{s}^{*}\Omega_{c}^{*}$ channels, respectively.
In the cases of the $PB(1/2^{+})$, $PB(3/2^{+})$, $VB(1/2^{+})$, and $VB(3/2^{+})$ sectors, which carry the quantum numbers $I(J^{P})=0(1/2^{-})$, $0(3/2^{-})$, $0(1/2^{-}, 3/2^{-})$, and $0(1/2^{-}, 3/2^{-}, 5/2^{-})$, respectively.
All these states have very small or zero widths, which are below the threshold of the most dominant channel about $20-30$ MeV. 
Some of these predicted molecular pentaquark candidates can be searched in the corresponding decay channels $\eta\Omega_{cc}^{(*)}$, $\phi\Omega_{cc}^{(*)}$, $\eta^{\prime}\Omega_{bc}^{\prime/*}$, $\phi\Omega_{bc}^{\prime/*}$, and so on.
The LHCb Collaboration is carrying out the relevant measurement and analysis, and these predicted states are likely to be detected in the near future.

\section*{Acknowledgements}

We would like to thank Fu-Lai Wang and Wen-Fei Wang for valuable discussions, and acknowledge Profs. Eulogio Oset and Kazem Azizi for careful reading the manuscript and useful comments. 
This work is supported by the China National Funds for Distinguished Young Scientists under Grant No. 11825503, the National Key Research and Development Program of China under Contract No. 2020YFA0406400, the 111 Project under Grant No. B20063, the fundamental Research Funds for the Central Universities under Grant No. lzujbky-2022-sp02, the project for top-notch innovative talents of Gansu province, and the National Natural Science Foundation of China (NSFC) under Grants No. 12247101, 12335001, 11965016, 11705069 and 12047501 (Z.F.S.). This work is 
also partly supported by the Natural Science Foundation of Changsha under Grants No. kq2208257, the Natural Science Foundation of Hunan province under Grant No. 2023JJ30647, the Natural Science Foundation of Guangxi province under Grant No. 2023JJA110076, and the NSFC under Grant No. 12365019 (C.W.X.).

 \addcontentsline{toc}{section}{References}
 

\begin{thebibliography}{9}

\bibitem{ParticleDataGroup:2022pth}
R.~L.~Workman \textit{et al.} [Particle Data Group],
PTEP \textbf{2022}, 083C01 (2022)

\bibitem{BaBar:2003oey}
B.~Aubert \textit{et al.} [BaBar],
Phys. Rev. Lett. \textbf{90}, 242001 (2003)
[arXiv:hep-ex/0304021 [hep-ex]].

\bibitem{CLEO:2003ggt}
D.~Besson \textit{et al.} [CLEO],
Phys. Rev. D \textbf{68}, 032002 (2003)
[erratum: Phys. Rev. D \textbf{75}, 119908 (2007)]
[arXiv:hep-ex/0305100 [hep-ex]].

\bibitem{Belle:2003nnu}
S.~K.~Choi \textit{et al.} [Belle],
Phys. Rev. Lett. \textbf{91}, 262001 (2003)
[arXiv:hep-ex/0309032 [hep-ex]].

\bibitem{LHCb:2015yax}
R.~Aaij \textit{et al.} [LHCb],
Phys. Rev. Lett. \textbf{115}, 072001 (2015)
[arXiv:1507.03414 [hep-ex]].

\bibitem{LHCb:2017iph}
R.~Aaij \textit{et al.} [LHCb],
Phys. Rev. Lett. \textbf{119}, no.11, 112001 (2017)
[arXiv:1707.01621 [hep-ex]].

\bibitem{LHCb:2017uwr}
R.~Aaij \textit{et al.} [LHCb],
Phys. Rev. Lett. \textbf{118}, no.18, 182001 (2017)
[arXiv:1703.04639 [hep-ex]].

\bibitem{Belle:2017ext}
J.~Yelton \textit{et al.} [Belle],
Phys. Rev. D \textbf{97}, no.5, 051102 (2018)
[arXiv:1711.07927 [hep-ex]].

\bibitem{LHCb:2021ptx}
R.~Aaij \textit{et al.} [LHCb],
Phys. Rev. D \textbf{104}, no.9, L091102 (2021)
[arXiv:2107.03419 [hep-ex]].

\bibitem{LHCb:2023rtu}
[LHCb],
[arXiv:2302.04733 [hep-ex]].

\bibitem{LHCb:2020tqd}
R.~Aaij \textit{et al.} [LHCb],
Phys. Rev. Lett. \textbf{124}, no.8, 082002 (2020)
[arXiv:2001.00851 [hep-ex]].

\bibitem{Matiunin:2020xbg}
V.~Matiunin [LHCb],
PoS \textbf{ICHEP2020}, 485 (2021)
[arXiv:2102.03175 [hep-ex]].

\bibitem{LHCb:2021xba}
R.~Aaij \textit{et al.} [LHCb],
Chin. Phys. C \textbf{45}, no.9, 093002 (2021)
[arXiv:2104.04759 [hep-ex]].

\bibitem{Liu:2013waa}
X.~Liu,
Chin. Sci. Bull. \textbf{59}, 3815-3830 (2014)
[arXiv:1312.7408 [hep-ph]].

\bibitem{Hosaka:2016pey}
A.~Hosaka, T.~Iijima, K.~Miyabayashi, Y.~Sakai and S.~Yasui,
PTEP \textbf{2016}, no.6, 062C01 (2016)
[arXiv:1603.09229 [hep-ph]].

\bibitem{Chen:2016qju}
H.~X.~Chen, W.~Chen, X.~Liu and S.~L.~Zhu,
Phys. Rept. \textbf{639}, 1-121 (2016)
[arXiv:1601.02092 [hep-ph]].

\bibitem{Richard:2016eis}
J.~M.~Richard,
Few Body Syst. \textbf{57}, no.12, 1185-1212 (2016)
[arXiv:1606.08593 [hep-ph]].

\bibitem{Lebed:2016hpi}
R.~F.~Lebed, R.~E.~Mitchell and E.~S.~Swanson,
Prog. Part. Nucl. Phys. \textbf{93}, 143-194 (2017)
[arXiv:1610.04528 [hep-ph]].

\bibitem{Olsen:2017bmm}
S.~L.~Olsen, T.~Skwarnicki and D.~Zieminska,
Rev. Mod. Phys. \textbf{90}, no.1, 015003 (2018)
[arXiv:1708.04012 [hep-ph]].

\bibitem{Guo:2017jvc}
F.~K.~Guo, C.~Hanhart, U.~G.~Mei\ss{}ner, Q.~Wang, Q.~Zhao and B.~S.~Zou,
Rev. Mod. Phys. \textbf{90}, no.1, 015004 (2018)
[erratum: Rev. Mod. Phys. \textbf{94}, no.2, 029901 (2022)]
[arXiv:1705.00141 [hep-ph]].

\bibitem{Liu:2019zoy}
Y.~R.~Liu, H.~X.~Chen, W.~Chen, X.~Liu and S.~L.~Zhu,
Prog. Part. Nucl. Phys. \textbf{107}, 237-320 (2019)
[arXiv:1903.11976 [hep-ph]].

\bibitem{Brambilla:2019esw}
N.~Brambilla, S.~Eidelman, C.~Hanhart, A.~Nefediev, C.~P.~Shen, C.~E.~Thomas, A.~Vairo and C.~Z.~Yuan,
Phys. Rept. \textbf{873}, 1-154 (2020)
[arXiv:1907.07583 [hep-ex]].

\bibitem{Meng:2022ozq}
L.~Meng, B.~Wang, G.~J.~Wang and S.~L.~Zhu,
Phys. Rept. \textbf{1019}, 1-149 (2023)
[arXiv:2204.08716 [hep-ph]].

\bibitem{Chen:2022asf}
H.~X.~Chen, W.~Chen, X.~Liu, Y.~R.~Liu and S.~L.~Zhu,
Rept. Prog. Phys. \textbf{86}, no.2, 026201 (2023)
[arXiv:2204.02649 [hep-ph]].

\bibitem{LHCb:2022ogu}
R.~Aaij \textit{et al.} [LHCb],
Phys. Rev. Lett. \textbf{131}, no.3, 031901 (2023)
[arXiv:2210.10346 [hep-ex]].

\bibitem{LHCb:2020jpq}
R.~Aaij \textit{et al.} [LHCb],
Sci. Bull. \textbf{66}, 1278-1287 (2021)
[arXiv:2012.10380 [hep-ex]].

\bibitem{LHCb:2019kea}
R.~Aaij \textit{et al.} [LHCb],
Phys. Rev. Lett. \textbf{122}, no.22, 222001 (2019)
[arXiv:1904.03947 [hep-ex]].

\bibitem{Wu:2010jy}
J.~J.~Wu, R.~Molina, E.~Oset and B.~S.~Zou,
Phys. Rev. Lett. \textbf{105}, 232001 (2010)
[arXiv:1007.0573 [nucl-th]].

\bibitem{Wang:2011rga}
W.~L.~Wang, F.~Huang, Z.~Y.~Zhang and B.~S.~Zou,
Phys. Rev. C \textbf{84}, 015203 (2011)
[arXiv:1101.0453 [nucl-th]].

\bibitem{Yang:2011wz}
Z.~C.~Yang, Z.~F.~Sun, J.~He, X.~Liu and S.~L.~Zhu,
Chin. Phys. C \textbf{36}, 6-13 (2012)
[arXiv:1105.2901 [hep-ph]].

\bibitem{Li:2014gra}
X.~Q.~Li and X.~Liu,
Eur. Phys. J. C \textbf{74}, no.12, 3198 (2014)
[arXiv:1409.3332 [hep-ph]].

\bibitem{Chen:2015loa}
R.~Chen, X.~Liu, X.~Q.~Li and S.~L.~Zhu,
Phys. Rev. Lett. \textbf{115}, no.13, 132002 (2015)
[arXiv:1507.03704 [hep-ph]].

\bibitem{Karliner:2015ina}
M.~Karliner and J.~L.~Rosner,
Phys. Rev. Lett. \textbf{115}, no.12, 122001 (2015)
[arXiv:1506.06386 [hep-ph]].

\bibitem{Chen:2015moa}
H.~X.~Chen, W.~Chen, X.~Liu, T.~G.~Steele and S.~L.~Zhu,
Phys. Rev. Lett. \textbf{115}, no.17, 172001 (2015)
[arXiv:1507.03717 [hep-ph]].

\bibitem{Ozdem:2018qeh}
U.~\"Ozdem and K.~Azizi,
Eur. Phys. J. C \textbf{78}, no.5, 379 (2018)
[arXiv:1803.06831 [hep-ph]].

\bibitem{Du:2019pij}
M.~L.~Du, V.~Baru, F.~K.~Guo, C.~Hanhart, U.-G.~Mei{\ss}ner, J.~A.~Oller and Q.~Wang,
Phys. Rev. Lett. \textbf{124}, no.7, 072001 (2020)
[arXiv:1910.11846 [hep-ph]].

\bibitem{Azizi:2022qll}
K.~Azizi, Y.~Sarac and H.~Sundu,
Phys. Rev. D \textbf{107}, no.1, 014023 (2023)
[arXiv:2210.03471 [hep-ph]].

\bibitem{LHCb:2021vvq}
R.~Aaij \textit{et al.} [LHCb],
Nature Phys. \textbf{18}, no.7, 751-754 (2022)
[arXiv:2109.01038 [hep-ex]].

\bibitem{Manohar:1992nd}
A.~V.~Manohar and M.~B.~Wise,
Nucl. Phys. B \textbf{399}, 17-33 (1993)
[arXiv:hep-ph/9212236 [hep-ph]].

\bibitem{Ericson:1993wy}
T.~E.~O.~Ericson and G.~Karl,
Phys. Lett. B \textbf{309}, 426-430 (1993)

\bibitem{Tornqvist:1993ng}
N.~A.~Tornqvist,
Z. Phys. C \textbf{61}, 525-537 (1994)
[arXiv:hep-ph/9310247 [hep-ph]].

\bibitem{Janc:2004qn}
D.~Janc and M.~Rosina,
Few Body Syst. \textbf{35}, 175-196 (2004)
[arXiv:hep-ph/0405208 [hep-ph]].

\bibitem{Ding:2009vj}
G.~J.~Ding, J.~F.~Liu and M.~L.~Yan,
Phys. Rev. D \textbf{79}, 054005 (2009)
[arXiv:0901.0426 [hep-ph]].

\bibitem{Molina:2010tx}
R.~Molina, T.~Branz and E.~Oset,
Phys. Rev. D \textbf{82}, 014010 (2010)
[arXiv:1005.0335 [hep-ph]].

\bibitem{Ohkoda:2012hv}
S.~Ohkoda, Y.~Yamaguchi, S.~Yasui, K.~Sudoh and A.~Hosaka,
Phys. Rev. D \textbf{86}, 034019 (2012)
[arXiv:1202.0760 [hep-ph]].

\bibitem{Li:2012ss}
N.~Li, Z.~F.~Sun, X.~Liu and S.~L.~Zhu,
Phys. Rev. D \textbf{88}, no.11, 114008 (2013)
[arXiv:1211.5007 [hep-ph]].

\bibitem{Xu:2017tsr}
H.~Xu, B.~Wang, Z.~W.~Liu and X.~Liu,
Phys. Rev. D \textbf{99}, no.1, 014027 (2019)
[erratum: Phys. Rev. D \textbf{104}, no.11, 119903 (2021)]
[arXiv:1708.06918 [hep-ph]].

\bibitem{Liu:2019stu}
M.~Z.~Liu, T.~W.~Wu, M.~Pavon Valderrama, J.~J.~Xie and L.~S.~Geng,
Phys. Rev. D \textbf{99}, no.9, 094018 (2019)
[arXiv:1902.03044 [hep-ph]].

\bibitem{Tang:2019nwv}
L.~Tang, B.~D.~Wan, K.~Maltman and C.~F.~Qiao,
Phys. Rev. D \textbf{101}, no.9, 094032 (2020)
[arXiv:1911.10951 [hep-ph]].

\bibitem{Ding:2020dio}
Z.~M.~Ding, H.~Y.~Jiang and J.~He,
Eur. Phys. J. C \textbf{80}, no.12, 1179 (2020)
[arXiv:2011.04980 [hep-ph]].

\bibitem{Hofmann:2005sw}
J.~Hofmann and M.~F.~M.~Lutz,
Nucl. Phys. A \textbf{763}, 90-139 (2005)
[arXiv:hep-ph/0507071 [hep-ph]].

\bibitem{Romanets:2012hm}
O.~Romanets, L.~Tolos, C.~Garcia-Recio, J.~Nieves, L.~L.~Salcedo and R.~G.~E.~Timmermans,
Phys. Rev. D \textbf{85}, 114032 (2012)
[arXiv:1202.2239 [hep-ph]].

\bibitem{Zhou:2018bkn}
Q.~S.~Zhou, K.~Chen, X.~Liu, Y.~R.~Liu and S.~L.~Zhu,
Phys. Rev. C \textbf{98}, no.4, 045204 (2018)
[arXiv:1801.04557 [hep-ph]].

\bibitem{Dong:2021bvy}
X.~K.~Dong, F.~K.~Guo and B.~S.~Zou,
Commun. Theor. Phys. \textbf{73}, no.12, 125201 (2021)
[arXiv:2108.02673 [hep-ph]].

\bibitem{Wang:2022aga}
W.~F.~Wang, A.~Feijoo, J.~Song and E.~Oset,
Phys. Rev. D \textbf{106}, no.11, 116004 (2022)
[arXiv:2208.14858 [hep-ph]].

\bibitem{Wang:2023aob}
F.~L.~Wang and X.~Liu,
Phys. Rev. D \textbf{108}, no.7, 074022 (2023)
[arXiv:2308.15255 [hep-ph]].

\bibitem{Kaiser:1995eg}
N.~Kaiser, P.~B.~Siegel and W.~Weise,
Nucl. Phys. A \textbf{594}, 325-345 (1995)
[arXiv:nucl-th/9505043 [nucl-th]].

\bibitem{Oller:1997ti}
J.~A.~Oller and E.~Oset,
Nucl. Phys. A \textbf{620}, 438-456 (1997)
[erratum: Nucl. Phys. A \textbf{652}, 407-409 (1999)]
[arXiv:hep-ph/9702314 [hep-ph]].

\bibitem{Oset:1997it}
E.~Oset and A.~Ramos,
Nucl. Phys. A \textbf{635}, 99-120 (1998)
[arXiv:nucl-th/9711022 [nucl-th]].

\bibitem{Kaiser:1998fi}
N.~Kaiser,
Eur. Phys. J. A \textbf{3}, 307-309 (1998).

\bibitem{Oller:1998hw}
J.~A.~Oller, E.~Oset and J.~R.~Pel\'aez,
Phys. Rev. D \textbf{59}, 074001 (1999)
[erratum: Phys. Rev. D \textbf{60}, 099906 (1999); erratum: Phys. Rev. D \textbf{75}, 099903 (2007)]
[arXiv:hep-ph/9804209 [hep-ph]].

\bibitem{Oller:2000ma}
J.~A.~Oller, E.~Oset and A.~Ramos,
Prog. Part. Nucl. Phys. \textbf{45}, 157-242 (2000)
[arXiv:hep-ph/0002193 [hep-ph]].

\bibitem{Oller:2000fj}
J.~A.~Oller and U.-G.~Mei{\ss}ner,
Phys. Lett. B \textbf{500}, 263-272 (2001)
[arXiv:hep-ph/0011146 [hep-ph]].

\bibitem{Wu:2010vk}
J.~J.~Wu, R.~Molina, E.~Oset and B.~S.~Zou,
Phys. Rev. C \textbf{84}, 015202 (2011)
[arXiv:1011.2399 [nucl-th]].

\bibitem{Xiao:2013yca}
C.~W.~Xiao, J.~Nieves and E.~Oset,
Phys. Rev. D \textbf{88}, 056012 (2013)
[arXiv:1304.5368 [hep-ph]].

\bibitem{Xiao:2019gjd}
C.~W.~Xiao, J.~Nieves and E.~Oset,
Phys. Lett. B \textbf{799}, 135051 (2019)
[arXiv:1906.09010 [hep-ph]].

\bibitem{Feijoo:2021ppq}
A.~Feijoo, W.~H.~Liang and E.~Oset,
Phys. Rev. D \textbf{104}, no.11, 114015 (2021)
[arXiv:2108.02730 [hep-ph]].

\bibitem{Dai:2023cyo}
L.~R.~Dai, L.~M.~Abreu, A.~Feijoo and E.~Oset,
[arXiv:2304.01870 [hep-ph]].

\bibitem{Debastiani:2017ewu}
V.~R.~Debastiani, J.~M.~Dias, W.~H.~Liang and E.~Oset,
Phys. Rev. D \textbf{97}, no.9, 094035 (2018)
[arXiv:1710.04231 [hep-ph]].

\bibitem{Liang:2017ejq}
W.~H.~Liang, J.~M.~Dias, V.~R.~Debastiani and E.~Oset,
Nucl. Phys. B \textbf{930}, 524-532 (2018)
[arXiv:1711.10623 [hep-ph]].

\bibitem{Liang:2020dxr}
W.~H.~Liang and E.~Oset,
Phys. Rev. D \textbf{101}, no.5, 054033 (2020)
[arXiv:2001.02929 [hep-ph]].

\bibitem{Dias:2018qhp}
J.~M.~Dias, V.~R.~Debastiani, J.~J.~Xie and E.~Oset,
Phys. Rev. D \textbf{98}, no.9, 094017 (2018)
[arXiv:1805.03286 [hep-ph]].

\bibitem{Yu:2018yxl}
Q.~X.~Yu, R.~Pavao, V.~R.~Debastiani and E.~Oset,
Eur. Phys. J. C \textbf{79}, no.2, 167 (2019)
[arXiv:1811.11738 [hep-ph]].

\bibitem{Yu:2019yfr}
Q.~X.~Yu, J.~M.~Dias, W.~H.~Liang and E.~Oset,
Eur. Phys. J. C \textbf{79}, no.12, 1025 (2019)
[arXiv:1909.13449 [hep-ph]].

\bibitem{Dias:2019klk}
J.~M.~Dias, Q.~X.~Yu, W.~H.~Liang, Z.~F.~Sun, J.~J.~Xie and E.~Oset,
Chin. Phys. C \textbf{44}, no.6, 064101 (2020)
[arXiv:1912.04517 [hep-ph]].


\bibitem{Bando:1984ej}
M.~Bando, T.~Kugo, S.~Uehara, K.~Yamawaki and T.~Yanagida,
Phys. Rev. Lett. \textbf{54}, 1215 (1985).
   
\bibitem{Bando:1987br}
M.~Bando, T.~Kugo and K.~Yamawaki,
Phys. Rept. \textbf{164}, 217-314 (1988).
  
\bibitem{Meissner:1987ge}
U.-G.~Mei{\ss}ner,
Phys. Rept. \textbf{161}, 213 (1988).

\bibitem{Mizutani:2006vq}
T.~Mizutani and A.~Ramos,
Phys. Rev. C \textbf{74}, 065201 (2006)
[arXiv:hep-ph/0607257 [hep-ph]].

\bibitem{Sakai:2017avl}
S.~Sakai, L.~Roca and E.~Oset,
Phys. Rev. D \textbf{96}, no.5, 054023 (2017)
[arXiv:1704.02196 [hep-ph]].

\bibitem{Close:1979}
F. E. Close, “An Introduction to Quarks and Partons”, Academic Press, Cambrige, 1979.

\bibitem{Oset:2001cn}
E.~Oset, A.~Ramos and C.~Bennhold,
Phys. Lett. B \textbf{527}, 99-105 (2002)
[erratum: Phys. Lett. B \textbf{530}, 260-260 (2002)]
[arXiv:nucl-th/0109006 [nucl-th]].

\bibitem{Jido:2003cb}
D.~Jido, J.~A.~Oller, E.~Oset, A.~Ramos and U.-G.~Mei{\ss}ner,
Nucl. Phys. A \textbf{725}, 181-200 (2003)
[arXiv:nucl-th/0303062 [nucl-th]].

\bibitem{Yamagata-Sekihara:2010kpd}
J.~Yamagata-Sekihara, J.~Nieves and E.~Oset,
Phys. Rev. D \textbf{83}, 014003 (2011)
[arXiv:1007.3923 [hep-ph]].

\bibitem{Yu:2022ymb}
G.~L.~Yu, Z.~Y.~Li, Z.~G.~Wang, J.~Lu and M.~Yan,
Nucl. Phys. B \textbf{990}, 116183 (2023)
[arXiv:2206.08128 [hep-ph]].

\bibitem{Marse-Valera:2022khy}
J.~A.~Mars\'e-Valera, V.~K.~Magas and A.~Ramos,
Phys. Rev. Lett. \textbf{130}, no.9, 9 (2023)
[arXiv:2210.02792 [hep-ph]].

\bibitem{Molina:2008jw}
R.~Molina, D.~Nicmorus and E.~Oset,
Phys. Rev. D \textbf{78}, 114018 (2008)
[arXiv:0809.2233 [hep-ph]].

\end{thebibliography}
\end{document}